\pdfoutput=1

\documentclass[final,1p,times,sort&compress]{elsarticle}

\usepackage{amssymb}
\usepackage{epsfig}
\usepackage{hyperref}
\usepackage{amsmath}

\begin{document}

\begin{frontmatter}



\title{single-quanta interferometry: which-way versus which-phase information stored in an ancillary quantum system}


\author[]{Soroush~Khademi\corref{cor1} }
\ead{soroush@physics.sharif.ir}
\author[]{Ali~Reza~Bahrampour}
\ead{bahrampour@sharif.edu}
\address{Department of Physics, Sharif University of Technology, P.O. Box 11155-9161, Tehran, Iran}

\cortext[cor1]{Corresponding author}

\begin{abstract}
In interferometers, the more information about the quanta's path available in an ancillary quantum system (AQS), the less visibility the interference has. By use of Shannon entropy, we try to compare the amount of which-phase information with the amount of which-way information stored in the AQS of two-path interferometers with symmetric beam merging. We show that the former is lower than or equal the latter if the bipartite system of the single-quanta and the AQS is initially prepared in a pure state and the interaction between the two parts is unitary. Especially when there exists symmetry, the equality holds. No which-way information is obtained by the measurement that we use for extracting the which-phase information and vice versa. In order to verify the results experimentally, we propose assembling a new single-photon interferometer.
\end{abstract}

\begin{keyword}
Complementarity\sep Wave-particle duality\sep Quantum erasure\sep Accessible information



\end{keyword}

\end{frontmatter}


\section{Introduction}
\label{S1}

There are two classical pictures of motion; a classical particle travels through one path and a classical wave can propagate through several paths. But in the quantum realm, a single quanta showing particle-like behavior in an interferometry experiment may exhibit wavelike behavior in another one. According to wave-particle complementarity, such a behavior change may happen only if the experimental setups are mutually exclusive \cite{falkenburg2007particle}; so the results of such experiments can be described by use of classical pictures of motion in a complementary way and without any contradiction \cite{home1997conceptual}. Many experiments support this principle, e.g. \cite{feynmann1970feynman,bogar1996entanglement}.

There exist interferometry setups in which the quanta's behavior is partially particle-like and partially wavelike. Sometimes, this partial behavior is observed when the quanta's path information is partially available in the environment, e.g. in an ancillary quantum system (AQS). The information can be extracted by an appropriate measurement. Based on the success probability of path attribution, a quantity called distinguishability (\textit{D}) is defined \cite{englert1996fringe} for measuring the particle-like behavior. Visibility (\textit{V}) of the interference quantifies the wavelike behavior. Englert \cite{englert1996fringe} proves that 

\begin{equation}
D^2 + V^2 \leq 1
\end{equation}
\label{d2v2}
\noindent
which means, as expected by wave-particle complementarity, a trade-off exists between the particle-like behavior and the wavelike behavior.

Different entropic measures are also employed for quantifying the partial behaviors or similar purposes. Such measures were used, first in 1979, by Wootters and Zurek \cite{wootters1979complementarity}. Other examples can be found in \cite{mittelstaedt1987unsharp,lahti1991some,kaszlikowski2003information,vaccaro2011particle,angelo2015wave}. These measures have been criticized by \cite{jaeger1995two} and \cite{durr2001quantitative}. Nevertheless, Coles et al. \cite{coles2014equivalence} took a major step in this way in 2014. They define a which-way guessing game and a which-phase guessing game played with a two-path interferometer. The possibility of winning each game represents one partial behavior and is determined by a special entropy. They finally find a very general relation, which is basically an entropic uncertainty relation, that shows the trade-off between partial behaviors.

Measuring the AQS in an inappropriate basis erases the which-way information, instead of extracting it. But, interestingly, such a measurement allows us to recognize sub-ensembles of the interferometry data showing more interference visibility. This phenomena is called quantum erasure \cite{scully1982quantum,scully1991quantum}. It has been observed in several experiments, e.g. \cite{durr1998origin,ma2013quantum}. The first attempt to quantify this phenomena was made by \cite{bjork1998complementarity}. After that, \cite{englert2000quantitative} derived an erasure relation setting an upper bound on the visibility of the sub-ensembles. The results of \cite{englert2000quantitative} are re-derived by \cite{coles2014equivalence} with the entropic approach.

We consider two guessing games like \cite{coles2014equivalence}; we use Shannon entropy as a standard information theoretic tool and try to compare the amount of which-way information stored in the AQS with the amount of which-phase information obtainable by using the AQS. We are also curious to know if extracting one of these two types of information is at the cost of losing the other one -- it can be expected by what is learnt from quantum erasure.

In Section (\ref{S2}), we determine our framework and elaborate the problem that we are going to deal with. The problem is solved for a special case with some interesting results in Section (\ref{S3}). We propose a new interferometry setup in Section (\ref{S4}) for experimental verification of the results. In Section (\ref{S5}), the strategies used in the games and the probable general answer of the problem are discussed. The conclusions are in Section (\ref{S6}).

\section{Framework}
\label{S2}

Figure 1.(a) shows a two-path interferometer. A qubit with Bloch vector $\textit{\textbf{q}}$ can represent the quanta's path;  the upper path and the lower one are characterized by $q_z=-1$ and $q_z=1$ respectively. The single quanta passes through a beam splitter (BS) represented by a single-qubit gate. It interacts with a phase shifter (PS). Gate $|0\rangle\langle0|+e^{i\phi}|1\rangle\langle1|$ plays the role of the PS. Then, it passes through a beam merger (BM). As we consider a symmetric beam merging for a single quanta, gate $exp(i\frac{\pi}{4}\sigma_y)$ can represent the BM. Finally, the quanta is detected in one of the paths -- the path qubit is measured in the computational basis.

In Figure 1.(b), we let an AQS interact with the quanta. Generally, there is no restriction on the AQS and the interaction; it can even change the path of the quanta. Correlations that may exist between the AQS and the quanta after the interaction make it possible to obtain some information about the quanta's path by using the AQS.

\begin{figure}[t!]
\centering
\includegraphics[scale=0.4]{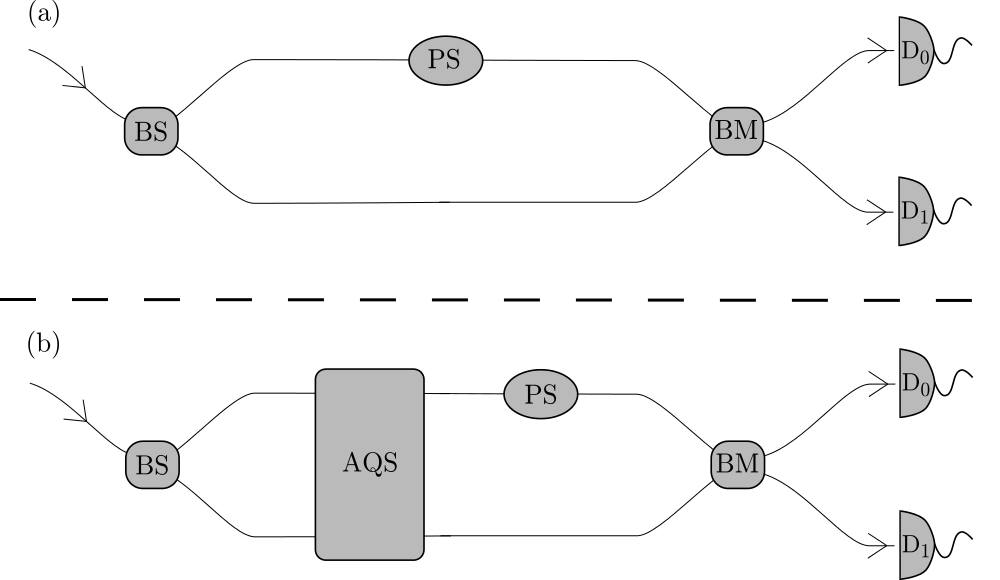}
\caption{(a) Scheme of an ordinary two-path interferometer. A beam splitter (BS) makes a linear combination of the two paths. Then a phase difference between the arms is made by a phase shifter (PS) and a beam merger (BM) combines the paths. Finally, one of the two detectors $D_0$ or $D_1$ detects the single quanta. (b) An ancillary quantum system (AQS) interacts with the single quanta within the interferometer. If the system is correlated with the quanta's spatial degree of freedom after this interaction, it stores some amount of information about the path of the single quanta.}
\label{FIG1}
\end{figure}

In order to find a value for the which-way information that may be stored in the AQS, let's consider a which-way guessing game (WWGG) played by Alice and Bob. While the BM is removed, Alice asks Bob to guess which detector will click or which one has clicked. Random variable $\cal W$ represents the detector that clicks. Bob can use the AQS. He performs a measurement on the AQS such that the amount of mutual information between $\cal W$ and the result of his measurement is maximized. Random variable $\cal M$ represents the result of such an optimum measurement. We take $I({\cal W} : {\cal M})$, which is the amount of accessible information for Bob, as the amount of which-way information stored in the AQS.

Let's come back to the closed interferometer. Alice and Bob play a which-phase guessing game (WPGG), like the one defined in \cite{coles2014equivalence}. One of the two values of random variable $\Phi = \{ \phi_0  ,  \phi_0+\pi \}$ is randomly applied to the interferometer as the phase difference by Alice. Bob should guess which value has been chosen. He can see the detectors and use the AQS. First of all, Bob looks at the detectors to see which one clicks. Random variable $\cal D$ represents the result. Then, given the value of $\phi_0$, he performs a measurement on the AQS which minimizes the amount of conditional Shannon entropy $H({\Phi} | {\cal D} , \text{the measurement's result})$. This measurement's result is represented by random variable $\cal E$. Alice's choice of $\phi_0$ has already maximized $H({\Phi} | {\cal D} , {\cal E})$ over all possible values of $\phi_0$. We take $H({\Phi} | {\cal D}) - H({\Phi} | {\cal D} , {\cal E})$ as the amount of which-phase information obtainable for Bob by using the AQS.

We want to know if there is a meaningful relationship between $I({\cal W} : {\cal M})$ and \linebreak $H({\Phi} | {\cal D}) - H({\Phi} | {\cal D} , {\cal E})$. Furthermore, we should survey how much information Bob gains if he uses the measurement of the WPGG in the WWGG and vice versa.

In the WWGG, we are faced with the problem of \textit{accessible information about quantum states} which is still, to the best of our knowledge, an open problem for the general case. In the WPGG, there are two quantum states; one of them is randomly prepared by Alice's choice between $\phi_0$ and $\phi_0+\pi$. Bob's measurement is supposed to maximize the value of
\begin{equation}
H(\Phi | {\cal D}) - H(\Phi | {\cal D} , {\cal E}) =  I(\Phi : {\cal D},{\cal E}) - I(\Phi : {\cal D}) =  I(\Phi : {\cal E}\:|\:{\cal D})
\end{equation}
\noindent
where the last term equals the expected value of mutual information between $\Phi$ and ${\cal E}$ given the value of ${\cal D}$. So, again, we are faced with the problem of accessible information. The problem has been solved \citep{levitin1995optimal} for two pure states. The accessible information about $|\psi^{(1)}\rangle$ and $|\psi^{(2)}\rangle$ with occurrence probability $p$ and $1-p$, respectively, is a function of $r=|\langle\psi^{(1)}|\psi^{(2)}\rangle|$ and $p$. We show this function with $I_A(r,p)$ -- see Appendix A for the exact form. As it is expected,
\begin{subequations}
\begin{eqnarray}
& I_{A} (r,p) \le I_{A} (r,0.5)\;, \label{iachar1} &\\
& I_{A} (r_1,p) \le I_{A} (r_2,p)\;\;\;\;\text{for}\;\;\;r_2 \le r_1\;. \label{iachar2} &
\end{eqnarray}
\end{subequations}
\noindent
In the next section, we will examine the situations that two pure states should be discriminated in the guessing games.

\section{Results}
\label{S3}

The spatial degree of freedom of the single quanta and the AQS form a bipartite system. We consider that the system's initial state is pure and the interaction between the two parts is unitary. After the interaction, the state of the system is of the form

\begin{equation}
|\Psi\rangle=\sqrt{p}\;|0\rangle \otimes |w_0\rangle + e^{i\gamma}\sqrt{1-p}\;|1\rangle \otimes |w_1\rangle
\end{equation}
\noindent
where $|w_0\rangle$ and $|w_1\rangle$ are two normalized states of the AQS; $0 \le p \le 1$ and $\gamma$ has been chosen such that $r_w=\langle w_0 | w_1 \rangle$ is a real non-negative number. By the PS, the system's state evolves into

\begin{equation}
|\Psi'\rangle=\sqrt{p}\;|0\rangle \otimes |w_0\rangle + e^{i(\gamma+\phi)}\sqrt{1-p}\;|1\rangle \otimes |w_1\rangle\;.
\end{equation}

Obviously, in the WWGG, Bob should discriminate $|w_0\rangle$ from $|w_1\rangle$ while the former exists with probability $p$. So

\begin{equation}
I({\cal W}:{\cal M}) = I_A (r_w,p)\;.
\label{RightSide}
\end{equation}

If the quanta passes through the BM, the system finds the state

\begin{equation}
|\Psi''\rangle = C_0\;|0\rangle \otimes |d_0^\phi\rangle - C_1\;|1\rangle \otimes |d_1^\phi\rangle
\label{final}
\end{equation}
\noindent
where

\begin{subequations}
\begin{eqnarray}
& C_j= \big( 0.5 + (-1)^j\;r\;\sqrt{p(1-p)}\cos(\gamma+\phi)\big)^{0.5}\;, \label{CJ} &\\
& |d_j^\phi\rangle = \dfrac{\sqrt{p}\;|w_0\rangle + (-1)^j e^{i(\gamma+\phi)}\sqrt{1-p}\;|w_1\rangle}{\sqrt{2}\;C_j}\;, \label{dinw} &\\
&& j=0,1\;. \nonumber
\end{eqnarray}
\end{subequations}

In the WPGG, if the upper (lower) detector clicks, Bob should discriminate $|d_{0}^{\phi_0}\rangle$ ( $|d_{1}^{\phi_0}\rangle$ ) from $|d_{0}^{\phi_0+\pi}\rangle$ ( $|d_{1}^{\phi_0+\pi}\rangle$ ). Since $|d_{0}^{\phi_0+\pi}\rangle=|d_{1}^{\phi_0}\rangle$ and $|d_{1}^{\phi_0+\pi}\rangle=|d_{0}^{\phi_0}\rangle$, $H(\Phi | {\cal D}) - H(\Phi | {\cal D} , {\cal E})$ equals the accessible information about two equiprobable states $|d_{0}^{\phi_0}\rangle$ and $|d_{1}^{\phi_0}\rangle$:

\begin{equation}
H(\Phi | {\cal D}) - H(\Phi | {\cal D} , {\cal E}) = I_A(r_d,0.5)
\label{LeftSide}
\end{equation}
\noindent
where

\begin{equation}
r_d=|\langle d_0^{\phi_0}|d_1^{\phi_0} \rangle | = \bigg( \dfrac{(2p-1)^2+4\;r_w^2\;p(1-p)\;\sin^2(\gamma+\phi_0)}{1-4\;r_w^2\;p(1-p)\;\cos^2(\gamma+\phi_0)}\bigg)^{0.5}
\end{equation}
\noindent
Now, we should fix the value of $\phi_0$. $r_d$ reaches its maximum at

\begin{equation}
\phi_0=\frac{\pi}{2}-\gamma\;.
\end{equation}
\noindent
So, based on (\ref{iachar2}) and (\ref{LeftSide}), the minimum of $H(\Phi | {\cal D}) - H(\Phi | {\cal D} , {\cal E})$ is at this value of $\phi_0$. Furthermore, by considering (\ref{final}) and (\ref{CJ}), it is easy to notify that $H(\Phi | {\cal D})$ reaches its maximum at the same value of $\phi_0$. As a conclusion, $H(\Phi | {\cal D} , {\cal E})$ is maximized by this value of $\phi_0$. Thus Alice plays the WPGG with this value of $\phi_0$ and $r_d$ equals $\sqrt{(2p-1)^2+4\;r_w^2\;p(1-p)}$\,.

For $p=0.5$, $r_d$ equals $r_w$ and, based on (\ref{RightSide}) and (\ref{LeftSide}), $H(\Phi | {\cal D}) - H(\Phi | {\cal D} , {\cal E}) = I({\cal W}:{\cal M})$. This is our most interesting result. It states that, in the case for which equality holds in (\ref{d2v2}), the amount of which-phase information obtainable by using the AQS equals the amount of which-way information obtainable by using it.

For an arbitrary $p$, $r_w \le r_d$. So it is not easy to compare the amount of $I_A(r_w,p)$ with the amount of $I_A(r_d,0.5)$ -- see (\ref{iachar1}) and (\ref{iachar2}). By explicit calculation of the value of these two functions for $0 \le r_{w} \le 1$ and $0 \le p \le 1$ and considering (\ref{RightSide}) and (\ref{LeftSide}), we have

\begin{equation}
H({\Phi} | {\cal D}) - H({\Phi} | {\cal D} , {\cal E}) \le I({\cal  W} : {\cal M})
\label{MR1}
\end{equation}
\noindent
where equality holds for $p=0, 0.5, 1$ or $r_w=1$. We note that the asymmetry makes it hard to reveal the value of the phase difference; as it does in an ordinary two-path interferometer (Figure 1.(a)).

The measurement found by \citep{levitin1995optimal} for extracting the accessible information about $|w_0\rangle$ and $|w_1\rangle$ is a projective measurement in the basis $\{|m_+\rangle,|m_-\rangle\}$ where

\begin{equation}
\begin{array}{rcl}
|w_{0}\rangle & = & \cos \frac{\alpha}{2}\;|m_+\rangle + \sin \frac{\alpha}{2}\;|m_-\rangle\;, \\
&&\\
|w_{1}\rangle & = & \cos \frac{\alpha'}{2}\;|m_+\rangle + \sin \frac{\alpha'}{2}\;|m_-\rangle\; ; \\
&&\\
\end{array}
\label{winm}
\end{equation}
\noindent
$\alpha$ and $\alpha'$ are functions of $p$ and $r_w$ and change from $0$ to $\pi$ -- see Appendix A. By considering (\ref{dinw}) with $\phi=\pi/2 - \gamma$ and (\ref{winm}), it is straightforward to show that $|\langle m_\pm | d_0^{\phi_0} \rangle|^2=|\langle m_\pm | d_1^{\phi_0} \rangle|^2$. So if Bob, in the WPGG, uses the measurement basis $\{|m_+\rangle,|m_-\rangle\}$, he gains no which-phase information.

Similarly, the measurement used for extracting the accessible information about $|d_{0}^{\phi_0}\rangle$ and $|d_{1}^{\phi_0}\rangle$ is a projective measurement in the basis $\{|e_+\rangle,|e_-\rangle\}$ where

\begin{equation}
\begin{array}{rcl}
e^{i\delta}\;|d_{0}^{\phi_0}\rangle & = & \cos \frac{\beta}{2}\;|e_+\rangle + \sin \frac{\beta}{2}\;|e_-\rangle\;, \\
&&\\
|d_{1}^{\phi_0}\rangle & = & \sin \frac{\beta}{2}\;|e_+\rangle + \cos \frac{\beta}{2}\;|e_-\rangle\; ; \\
&&\\
\end{array}
\label{dine}
\end{equation}
\noindent
$\delta$ is the argument of the complex number $\langle d_0^{\phi_0}|d_1^{\phi_0} \rangle$ and $\beta$ equals $arcsin(r_d)$ -- see Appendix A. By considering (\ref{dinw}) with $\phi=\pi/2 - \gamma$ and (\ref{dine}), it can be shown that $|\langle e_\pm | w_0 \rangle|^2=|\langle e_\pm | w_1 \rangle|^2$. So if Bob, in the WWGG, measures the AQS in the basis $\{|e_+\rangle,|e_-\rangle\}$, he gains no which-way information. The measurement used in the WPGG is an erasing measurement.

\section{Experimental verification}
\label{S4}

\begin{figure}[t!] 
\centering
\includegraphics[scale=0.4]{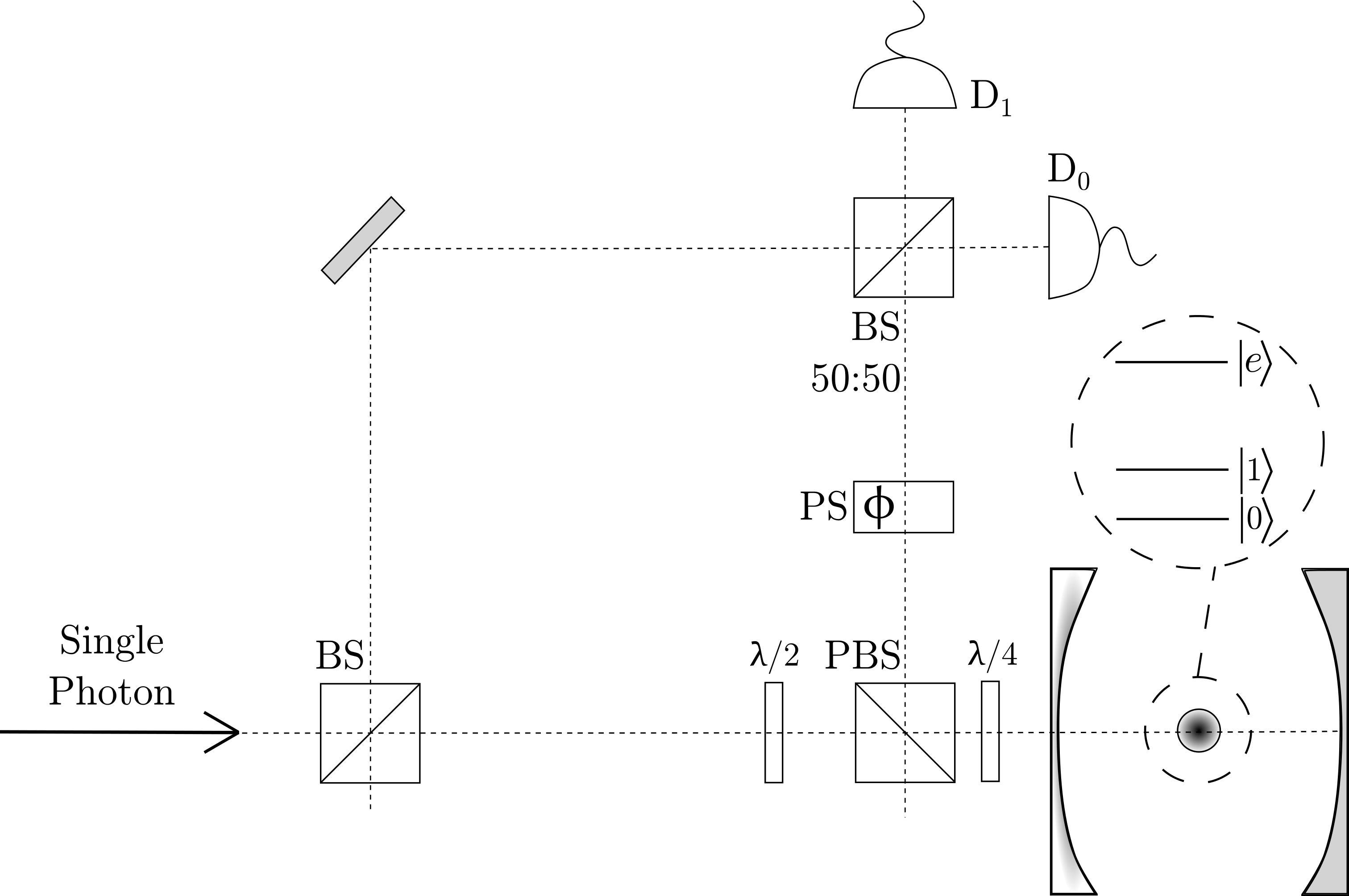}
\caption{A single photon is sent toward a beam splitter (BS). In the lower arm, it interacts with a three-level atom trapped in a high-finesse cavity -- the resonance frequency of the cavity equals the transition frequency between $|1\rangle$ and $|e\rangle$ and is around the frequency of the single photon. The polarization beam splitter (PBS) and the wave plates are used for directing the photon in the path. Then the photon passes through a phase shifter (PS) and experiences a $\phi$ phase shift. A 50:50 BS merges the two arms. Finally, $D_0$ or $D_1$ detects the single photon.}
\label{FIG2}
\end{figure}

In order to check relation (\ref{MR1}) experimentally, we propose assembling the interferometer depicted in Figure (\ref{FIG2}). It's a modified Mach-Zehnder interferometer. A single photon with linear polarization $\cal V$ enters the interferometer through a BS. In the lower arm, a half-wave plate changes its polarization to $\cal H$ (perpendicular to $\cal V$) and the polarization beam splitter (PBS) lets it pass. Its polarization is changed to circular by a quarter-wave plate. Then it interacts with an atom trapped in a cavity (as an AQS); after being reflected, it passes through the quarter-wave plate for the second time. Since the photon's polarization is changed to $\cal V$, the PBS reflects it this time. An adjustable PS exerts a $\phi$ phase shift; we note that the optical path lengths of the two arms have been equalized before inserting the PS. Then a 50:50 BS plays the role of a BM. Finally, $D_0$ or $D_1$ detects the single photon.

In order to describe the AQS, let's consider an atom with three levels $\{ |0\rangle, |1\rangle, |e\rangle \}$ trapped in a cavity. The transition frequency between $|1\rangle$ and $|e\rangle$ equals the resonance frequency of the cavity but the transition frequencies between $|0\rangle$ and $|1\rangle$ and between $|0\rangle$ and $|e\rangle$ are far from it. By making the cavity's left reflector have a little leakage, the system can interact with a (circularly polarized) ultra-narrow-band single photon coming from the left side. The photon is finally reflected. Due to this interaction, the atomic state $|+\rangle=(|1\rangle+|0\rangle)/\sqrt{2}$ is transformed to $(|1\rangle+e^{i\eta}|0\rangle)/\sqrt{2}$ where $\eta$ can be set by adjusting the central frequency of the incoming photon -- for more details, see Appendix B. This system has been implemented very well \cite{reiserer2013nondestructive} -- it has been proposed for several applications, e.g. \cite{duan2004scalable,reiserer2013nondestructive,cho2005generation}.

In the interferometer, we prepare the atom in $|+\rangle$ and send the single photon. Let's consider the situation in which the first BS is symmetric ($p=0.5$) and $\eta=\pi/2$. In comparison with Section (\ref{S3}), $|w_0\rangle=|+\rangle$ and $|w_1\rangle=exp(-i\pi/4)(|1\rangle+i|0\rangle)/\sqrt{2}$.

In order to play the WWGG with this setup, the second BS is discarded; the atom is prepared in $|+\rangle$ and the single photon is sent. Bob should guess which detector will click. He uses the AQS by performing a measurement on the atom in the optimum basis $\big\{(e^{i\pi/8}|1\rangle+e^{-i\pi/8}|0\rangle)/\sqrt{2}, (e^{-i3\pi/8}|1\rangle+e^{i3\pi/8}|0\rangle)/\sqrt{2}\big\}$. Alice and Bob repeat this procedure many times and estimate $I({\cal W} : {\cal M})$.

In the WPGG, Alice randomly adjusts the PS such that $\phi=3\pi/4$ or $\phi=7\pi/4$ and runs the prepared setup -- she is playing with the cleverest choice of $\phi_0$. Bob should make a guess about Alice's choice. He sees which detector has clicked and measures the energy level of the atom, i.e. in the basis $\{|0\rangle, |1\rangle\}$. They repeat this procedure again and again and estimate $H({\Phi} | {\cal D}) - H({\Phi} | {\cal D} , {\cal E})$.

The two estimated amounts of information calculated based on the results of these guessing games will be approximately equal.

\section{Discussion}
\label{S5}

In the WPGG, Alice does her best to make the game hard. In both games, Bob does his best to win the games. We should keep in mind that their strategies are based on maximization or minimization of $H({\Phi} | {\cal D} , {\cal E})$. Nevertheless, in the cases that we examined, if their strategies were based on Bob's \textit{mean error probability}, they would have the same choices for the value of $\phi_0$ and the bases of the measurements \cite{helstrom1976quantum,helstrom1967detection}.

The problem of accessible information has been also solved for some other special situations including the one of discriminating between two mixed states of a qubit that have the same determinant \cite{levitin1995optimal,suzuki2007accessible}. It allows us to investigate the validity of (\ref{MR1}) in some special examples that Bob should discriminate between two mixed states, e.g. when the initial state of the single-quanta or the AQS is not pure. Based on our results in Section (\ref{S3}) and such examples, we conjecture that relation (\ref{MR1}) holds for the general case of all binary interferometers -- such interferometers have just two interfering paths \cite{coles2014equivalence} -- with symmetric beam merging. 
 
\section{Conclusions}
\label{S6}

By use of Shannon entropy, we set the measures of which-phase and which-way information that are achievable through the AQS of two-path interferometers with symmetric beam merging. We investigated their relationship for the situations in which the bipartite system of the single-quanta and the AQS was prepared in a pure state and the interaction between the two parts was unitary. It was shown that the obtainable amount of which-phase information is lower than or equal the amount of which-way information. For the symmetric case, the equality holds -- this is the case for which there exists a sharp trade-off between distinguishability and visibility. The measurement which we used for extracting the which-phase information erases the whole amount of which-way information and vice versa. We also proposed a setup for experimental verification of the results; it is feasible by today's technology.

\section*{Acknowledgements}

The authors thank Vahid Karimipour, Farnaz Farman and Marzieh Bathaee for very helpful discussions.

\section*{Appendix A: The explicit form of $I_A(r,p)$ and the measurement used for extracting the information}

In this appendix, we review some results of reference \citep{levitin1995optimal}. The accessible information about two pure quantum states with density matrices $\rho^{(1)}=|\psi^{(1)}\rangle\langle\psi^{(1)}|$ and $\rho^{(2)}=|\psi^{(2)}\rangle\langle\psi^{(2)}|$, that occur respectively with probability $p^{(1)}=p$ and $p^{(2)}=1-p$, is obtained by a projective measurement in the basis $\{|\varphi^{(1)}\rangle,|\varphi^{(2)}\rangle\}$; in this basis,

\begin{equation}
p \rho^{(1)}_{nn'} = (1-p)  \rho^{(1)}_{nn'} \;\;\;\;\;\; \text{for} \;\, n \neq n' \;.
\end{equation}
\noindent
So, if we write

\begin{equation}
\rho^{(j)}=\dfrac{I+\textit{\textbf{s}}^{(j)}.\boldsymbol{\sigma}}{2}\;\;\;\;\;\, j=1,2\;,
\end{equation}
\noindent
we have

\begin{equation}
p\,(s^{(1)}_x\mp is^{(1)}_y) = (1-p)\,(s^{(2)}_x\mp is^{(2)}_y)\;.
\end{equation}
\noindent
We are allowed to consider $r=\langle\psi^{(1)}|\psi^{(2)}\rangle$ as a real non-negative number and set $s^{(1)}_y=s^{(2)}_y=0$. By considering $ tr(\rho^{(1)}\rho^{(2)})=r^2$ and the unity of Bloch vectors $\textit{\textbf{s}}^{(1)}$ and $\textit{\textbf{s}}^{(2)}$, it is straightforward to find $s^{(j)}_x$ and $s^{(j)}_z$ for $j=1,2$.

In order to find the measurement basis for $p=0.5$, we note that for this value of $p$

 \begin{equation}
 \begin{array}{rcl}
 & s^{(1)}_x=s^{(2)}_x=r\:, &\\
 \\
 & s^{(1)}_z=-s^{(2)}_z=\sqrt{1-r^2}\: . &\\
 \end{array}
 \end{equation}
 \noindent
So we can write

\begin{equation}
\begin{array}{rcl}
& |\psi^{(1)}\rangle = \cos \dfrac{\chi}{2}\;|\varphi^{(1)}\rangle + \sin \dfrac{\chi}{2}\;|\varphi^{(2)}\rangle\;, &\\
&&\\
& |\psi^{(2)}\rangle = \sin \dfrac{\chi}{2}\;|\varphi^{(1)}\rangle + \cos \dfrac{\chi}{2} |\varphi^{(2)}\rangle &\\
&&\\
\end{array}
\end{equation}
\noindent
where $\chi$ equals $arcsin(r)$. For an arbitrary $p$, we have

\begin{equation}
\begin{array}{rcl}
& |\psi^{(j)}\rangle = \cos \dfrac{\theta^{(j)}}{2}\;|\varphi^{(1)}\rangle + \sin \dfrac{\theta^{(j)}}{2}\;|\varphi^{(2)}\rangle\;, &\\
&&\\
& \theta^{(j)} = arccos(s_z^{(j)})\;, &\\
&&\\
&& j=1,2\;. \\
&&\\
\end{array}
\end{equation}

In order to find the value of accessible information, we note that

\begin{equation}
I_A(r,p) = -\displaystyle\sum_{n=1}^2 \rho_{nn} \log_2 \rho_{nn}\: + \:\displaystyle\sum_{j=1}^2\displaystyle\sum_{n=1}^2 p^{(j)}\,\rho_{nn}^{(j)} \log_2 \rho_{nn}^{(j)}
\end{equation}
\noindent
where $\rho=p\,\rho^{(1)}+(1-p)\,\rho^{(2)}$. By defining $C=\sqrt{1-4p(1-p)r^2}$ and using the value of $s^{(1)}_z$ and $s^{(2)}_z$, it can be shown that
\begin{equation}
\begin{array}{rcl}
I_A(r,p)=&\big(1/2C\big)\;\bigg\{p\Big[\big(C+1-2(1-p)r^2\big)\log_2\big(C+1-2(1-p)r^2\big)\;\;\;\;\;\;\;\;\;\;\;\;\;\;&\\
&&\\
&+\big(C-1+2(1-p)r^2\big)\log_2\big(C-1+2(1-p)r^2\big)\Big]\;\;\;\;\;\;\;\;\;\;\;\;\;\;\;\;\;\;\;\;\;\;\;\;\;\;\;&\\
&&\\
&+(1-p)\Big[\big(C-1+2pr^2\big)\log_2\big(C-1+2pr^2\big)\;\;\;\;\;\;\;\;\;\;\;\;\;\;\;\;\;\;\;\;\;\;\;\;\;\;\;\;\;\;\;\;\;\;&\\
&&\\
&+\big(C+1-2pr^2\big)\log_2\big(C+1-2pr^2\big)\Big]\;\;\;\;\;\;\;\;\;\;\;\;\;\;\;\;\;\;\;\;\;\;\;\;\;\;\;\;\;\;\;\;\;\;\;\;\;\;\;\;\;\;\;\;&\\
&&\\
& -\big(C+1-2p\big)\log_2\big(C+1-2p\big)-\big(C-1+2p\big)\log_2\big(C-1+2p\big) \bigg\}\;.&\\
&&\\
\end{array}
\end{equation}

\section*{Appendix B: The AQS of the proposed setup}

In order to describe the AQS, let's consider an atom with three levels $\{ |0\rangle, |1\rangle, |e\rangle \}$ trapped in a cavity. The transition frequency between $|1\rangle$ and $|e\rangle$ equals the resonance frequency of the cavity ($f_0$) but the transition frequencies between $|0\rangle$ and $|1\rangle$ and between $|0\rangle$ and $|e\rangle$ are far from it. When the atom's state is in the subspace spanned by $|1\rangle$ and $|e\rangle$, strong coupling between the atom and the cavity leads to mode splitting; that is to say, the system finds entangled energy levels and the transition frequency between these new levels is different from $f_0$. Now, we make the cavity's left reflector have a little leakage and let the system interact with a (circularly polarized) ultra-narrow-band single photon coming from the left side. What happens depends on the atom's initial state and the photon's frequency.

To be accurate, we use the relation between input and output operators of this system derived by \cite{duan2004scalable}:

\begin{equation}
\hat{a}_{out} \approx \frac{i \Delta - \kappa / 2}{i\Delta + \kappa / 2} \, \hat{a}_{in}
\end{equation}
\noindent
where $\kappa$ is the cavity decay rate and

\begin{equation}
\Delta= 2\pi (f_p - f_s)\: ;
\end{equation}
\noindent
$f_p$ is the frequency of the single photon and $f_s$ is the resonance frequency of the system which the photon faces. If the atom is initially prepared in $|0\rangle$, it is not coupled with the cavity; the photon faces just the cavity and $f_s = f_0$. If the atom is initially prepared in the subspace spanned by $|1\rangle$ and $|e\rangle$, it is strongly coupled with the cavity; $f_s$ equals the transition frequency between the entangled energy levels of the atom and the cavity.

We are curious about the situation in which $f_p$ is around $f_0$. When the atom is prepared in $|0\rangle$, $\Delta$ is comparable with $\kappa$ and the photon experiences an $\eta \approx \pi - 2\,arctan(2\Delta/\kappa)$ phase shift by being reflected (for $\Delta \ge 0$). But when the atom is prepared in the subspace spanned by $|1\rangle$ and $|e\rangle$,  $\Delta \gg \kappa$ and the photon does not experience a phase shift after being reflected. The atomic state $(|1\rangle+|0\rangle)/\sqrt{2}$ evolves into  $(|1\rangle+e^{i\eta}|0\rangle)/\sqrt{2}$ by the interaction.


\section*{References}
\bibliographystyle{elsarticle-num}




\end{document}